\documentclass[twocolumn,superscriptaddress,prl]{revtex4-1}
\usepackage{amsmath,amssymb}
\usepackage{graphicx}
\usepackage{color}
\usepackage[colorlinks,bookmarks=false,citecolor=blue,linkcolor=red,urlcolor=blue]{hyperref}
\usepackage{times}

\def \trm{\textrm}

\def \hf{\tfrac{1}{2}}    
\def \u{\uparrow}
\def \d{\downarrow}

\def \ord{\mathcal{O}}

\newcommand{\bra}[1]{\langle\left.{#1}\right|}
\newcommand{\ket}[1]{\left|{#1}\right.\rangle}

\newcommand{\NN}{\mathcal{N}} 

\begin{document}

\author{Vincenzo Alba}
\author{Masudul Haque}
\affiliation{Max-Planck-Institut f\"{u}r Physik komplexer Systeme, N\"{o}thnitzer Stra{\ss}e 38, D-01187 Dresden, Germany}
\author{Andreas~M.~L\"auchli}
\affiliation{Max-Planck-Institut f\"{u}r Physik komplexer Systeme, N\"{o}thnitzer Stra{\ss}e 38, D-01187 Dresden, Germany}
\affiliation{Institute f\"ur Theoretische Physik, Universit\"at Innsbruck, A-6020 Innsbruck, Austria}

\date{\today}

\title{Boundary-locality and perturbative structure of entanglement spectra in
  gapped systems}

\begin{abstract} 

The entanglement between two parts of a many-body system can be characterized
in detail by the entanglement spectrum.  Focusing on gapped phases of several
one-dimensional systems, we show how this spectrum is dominated by
contributions from the boundary between the parts.  This contradicts the view
of an ``entanglement Hamiltonian'' as a bulk entity.  The boundary-local
nature of the entanglement spectrum is clarified through its hierarchical
level structure, through the combination of two single-boundary spectra to
form a two-boundary spectrum, and finally through consideration of dominant
eigenfunctions of the entanglement Hamiltonian.  We show consequences of
boundary-locality for perturbative calculations of the entanglement spectrum.

\end{abstract}


\maketitle

\paragraph*{Introduction ---}

The study of entanglement-related quantities in condensed matter has led to a
large body of interdisciplinary work \cite{AmicoFazioOsterlohVedral_RMP08}.
Recently, the concept of the \emph{entanglement spectrum} has established itself at the
forefront of the field, as this spectrum provides much finer information than
a single number like the von Neumann entropy.
Considering a bipartition of the system into parts $A$ and $B$, the
entanglement spectrum (ES), $\{\xi_i\}$, is defined in terms of the Schmidt
decomposition
\begin{equation}  \label{eq:schmidt_ES_defn}
\ket{\psi}=\sum_i e^{-\xi_i/2}\ket{\psi_i^A}\otimes\ket{\psi_i^B}. 
\end{equation}
Here $\ket{\psi}$ is the ground state, and the states $\ket{\psi_i^A}$
($\ket{\psi_i^B}$) form an orthonormal basis for the subsystem $A$ ($B$).  The
ES $\{\xi_i=-\log\lambda_i\}$ can also be thought of in terms of the
eigenvalues $\{\lambda_i\}$ of the reduced density matrix $\rho_A$ obtained
after tracing out the $B$ part of the system density matrix
$\ket{\psi}\bra{\psi}$.

The ES has been studied earlier for insights into 
the density matrix renormalization group (DMRG) algorithm \cite{Okunishi-99,
  PeschelEisler_JPA09, ChungPeschel_PRB01}, and more recently because of its
relation to low-energy boundary-related modes in topological phases
\cite{topological_ES_various,Laeuchli_PRL10,LiHaldane_PRL08}.  
One way of thinking of the ES is in terms of an \emph{entanglement
  Hamiltonian} (EH) $H_A=-\log\rho_A$ acting only on the $A$ degrees of freedom
\cite{Nienhuis-09, PeschelEisler_JPA09, Poilblanc-10, Schliemann_2011,Peschel11}.  The
ES is then the spectrum of this object.  This gives a suggestive
correspondence to the energy spectrum of true bulk Hamiltonians.  This idea
has also inspired calculations of bulk terms that appear in such a
`Hamiltonian' \cite{Nienhuis-09, PeschelEisler_JPA09}.  In topologically
ordered systems such as fractional quantum Hall states, the EH indeed seems to
share the low lying edge modes with the physical Hamiltonian of a system with
the same boundary.  This correspondence gave rise to the loose idea that
entanglement Hamiltonians are quite similar to physical "bulk" Hamiltonians,
apart perhaps from non-essential renormalizations, like longer range
hoppings or spatial inhomogeneities \cite{LiHaldane_PRL08,Laeuchli_PRL10}.
However, the well-known \emph{area law} for the entanglement entropy
$S_A=-\sum\lambda_i\log\lambda_i$, stating that $S_A$ scales with the size of
the boundary between $A$ and $B$ parts, clearly suggests that the spectrum of 
$\rho_A$ (and hence also the ES and EH) must in some sense be dominated 
by the boundary degrees of freedom.

This contradiction is highlighted even more by considering states that are 
described by simple matrix product states, such as the Majumdar-Ghosh point in the
frustrated antiferromagnetic chain ($J_1$-$J_2$ model) or the AKLT (Affleck-Kennedy-
Lieb-Tasaki) state
\cite{PerezGarciaVerstraeteWolfCirac_QIC07}. The entanglement spectrum for
these states consists only of small number of finite values, independent of
the (sufficiently large) block size A.  These simple gapped states thus
clearly have entanglement spectra determined completely by the boundary and
not by the bulk.

\begin{figure*}
\includegraphics*[width=\linewidth]{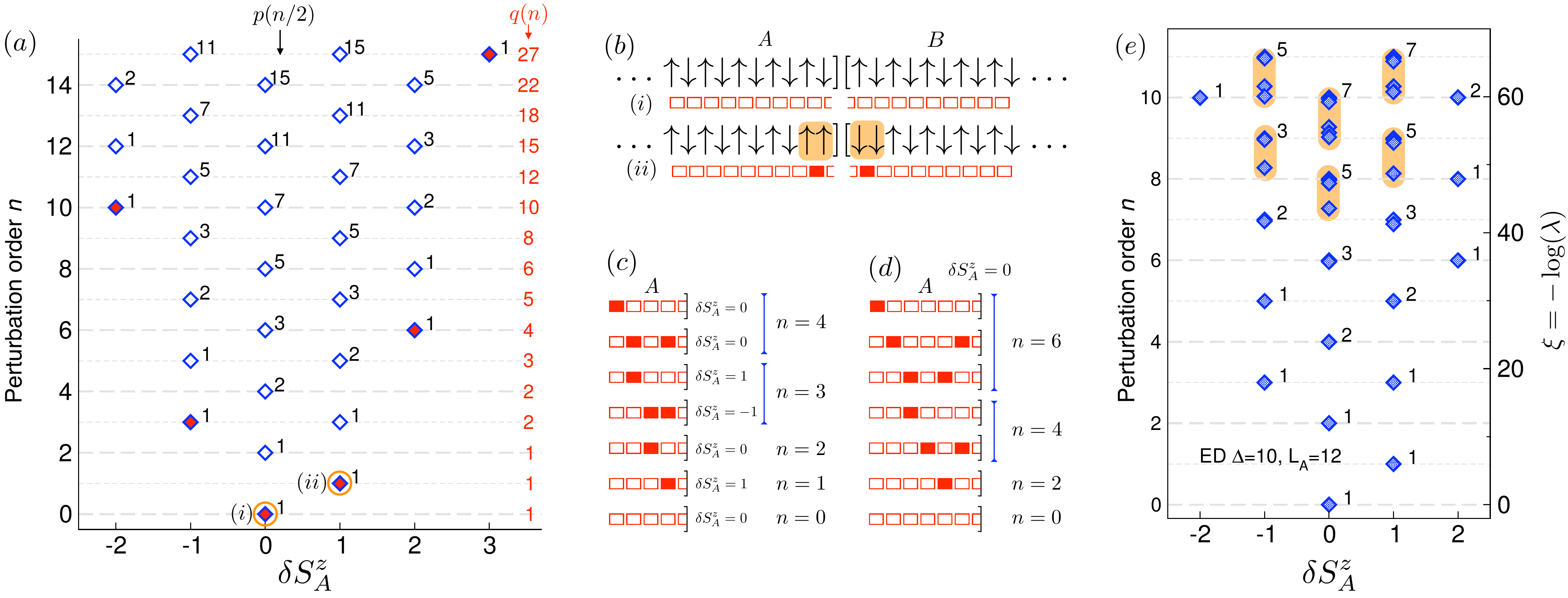}
\caption{(Color online) Single boundary entanglement spectrum (ES) of the XXZ chain. (a) Long
  or infinite blocks.  The ES levels appear at well-defined levels, which
  correspond to successive perturbative orders $n$.  The accompanying integers
  denote degeneracies.  On the right we show the total degeneracies at each
  order.  The levels marked (i) and (ii) have dominant contributions shown in
  (b), where a filled-rectangle notation is introduced for domain walls.  (c)
  Dominant contributions to low-lying levels.  (d) Dominant contributions to
  low-lying ${\delta}S_A^z=0$ levels. (e) Finite-size block (12 spins in $A$);
  arbitrary-precision exact diagonalization data for $\Delta=10$.  The effects
  of finite size are seen through breaking of degeneracies at order $\sim{8}$
  and higher.  The integer-indicated degeneracies are now approximate.  }
\label{fig:cartoon}
\end{figure*}

In this work, we sharpen the boundary picture of the entanglement spectrum by
considering more general gapped one-dimensional (1D) systems.  The
boundary-local nature of the ES is demonstrated and made quantitative in
several ways.  We show that the ES of gapped states generally have a
hierarchical structure, with excitations farther from the boundary being
successively higher in the ES.  This also leads us to the idea that the ES can
be obtained from a boundary-connected perturbative calculation, and we show
elements of such a calculation.  We also show that the ES of two-boundary
blocks can be constructed by combining the ES of single-boundary blocks,
demonstrating that the bulk degrees of freedom play a secondary role.  We will
concentrate on the widely familiar XXZ chain, and as a bonus, we uncover a
beautiful set of degeneracy structures in the ES of this model.  
We believe these findings are very general; we emphasize this by briefly
presenting boundary locality in two other systems, namely the Heisenberg
ladder and the Mott insulating phase of the Bose-Hubbard model.

In our examples we consider perturbations around a product state; it should be
possible to adapt to perturbations around a point which is a matrix product
state with small-rank matrices.  Gapped 1D phases often contain or are
connected to such simple points.
Gapped phases in general are not very universal in their low-energy
properties; it is therefore remarkable that the ES has features common in many
gapped phases, which we identify in this work.

The usual methods for calculating ES are numerical exact diagonalization or
DMRG, or using two-point correlators for non-interacting systems
\cite{ChungPeschel_PRB01}.  We propose our boundary-linked perturbation theory
as a general alternate technique for calculating ES in gapped phases
containing a simple point around which one can perturb.
The basic insight is that, while calculating the complete reduced density matrix 
perturbatively requires perturbation terms acting on the whole system/block,
if one is only interested in determining the ES and the leading order eigenfunctions
(\emph{entanglement eigenstates}), then a boundary local perturbation theory is 
able to construct the ES order by order in a physically transparent way.

\paragraph*{Boundary-linked perturbation theory  ---}
To obtain ES levels correctly up to $n$-th order, one only needs to consider
perturbations up to this order which act within ${\alpha}n$ sites from the
boundary, $\alpha$ depending on the type of perturbation.  This is a direct
but powerful consequence of boundary-locality, namely, that the ES levels at
increasing orders correspond to increasing distances from the boundary.  The
perturbative calculation necessary for calculating ES levels is thus
``boundary-linked''.

For any two orthonormal bases $|\varphi_i^A\rangle, |\varphi_j^B\rangle$, of
subsystems $A$ and $B$, the system wavefunction can be written as
$|\psi\rangle=\sum\limits_{i,j}\textbf{M}_{ij}|\varphi^A_i\rangle\otimes|\varphi^B_j\rangle$.
The $e^{-\xi_i/2}$ of Eq.\ \eqref{eq:schmidt_ES_defn} are obtained from a
singular value decomposition (SVD) of the matrix ${\bf M}$.  Further aspects
of the perturbative calculation of ES can be formulated in terms of the matrix
${\bf M}$.  For example, when calculating ${\bf M}$ perturbatively, we find
that a contribution to ${\bf M}$ at some order can lead to a new ES level only
if the contribution is not appearing on the same row or the same column as a
previous contribution which led to a new ES level at lower order.  In physical
terms, this means that we do not get new ES levels by applying a perturbation
only to the $A$ block but need to apply perturbation terms to both blocks in a
linked way.
Some perturbative calculations for the Heisenberg chain, and some general
perturbative features of the matrix ${\bf M}$, appear in the Supplementary
Materials.

\paragraph*{The XXZ gapped phase ---}

We consider the anisotropic Heisenberg (XXZ) chain, ${\mathcal H}\equiv
{\mathcal H_z}+{\mathcal H_{xy}}$, with ${\mathcal
  H_z}\equiv\Delta\sum\limits_iS^z_iS^z_{i+1}$, and ${\mathcal H_{xy}}\equiv
\tfrac{1}{2}\sum\limits_i(S_i^+S_{i+1}^- +h.c.)$, in the gapped phase
$\Delta>1$.  The Ising limit $\Delta\gg1$ is simple: the ground state is
spanned by the two product (N\'eel) states $|\trm{N1}\rangle\equiv
|\uparrow\downarrow \uparrow\downarrow\uparrow\downarrow\cdots\rangle$ and
$|\trm{N2}\rangle\equiv |\downarrow\uparrow\downarrow\uparrow
\downarrow\uparrow\cdots\rangle$.  We will therefore think perturbatively
around this simple limit, ${\mathcal H_{xy}}$ being the perturbation.
In the language of `domain walls' (bond between neighboring aligned spins),
the perturbation ${\mathcal H}_{xy}$ can have two effects: it can create a
pair of domain walls situated two bonds apart, or it can move a domain wall
over two sites.
We choose for simplicity to work with perturbations around a single N\'eel
state, by selecting $\ket{\textrm{N1}}$ instead of the linear combination
$\tfrac{1}{\sqrt{2}}(\ket{\textrm{N1}}+\ket{\textrm{N2}})$ to be the vacuum state.
For cases where the linear combination is appropriate, the ES can readily be
reconstructed from knowledge about the individual N\'eel states by overlaying
and shifting by $\log 2$ of the single state ES.

\paragraph*{The XXZ  single-boundary ES ---}

Fig.~\ref{fig:cartoon}(a) shows the ES for the simplest setup, namely, a
single boundary partitioning an open XXZ chain into two long blocks. The total
$S^z$ of the block $A$ is a good quantum number for the reduced density matrix
$\rho_A$.  Hence we organize the ES into sectors of $\delta{S_A^z}$, which is
the difference of the $S_A^z$ value of the corresponding entanglement
eigenstate from the block $S^z$ value in the N\'eel state (0 or
$\pm\tfrac{1}{2}$).
The most notable feature is that the ES is organized in equally spaced
hierarchical levels.
In Figs.\ \ref{fig:cartoon}(a) and \ref{fig:cartoon}(e) the hierarchical
orders are denoted as ``perturbation order'' because successive orders in
a perturbative calculation yield successively higher ES levels. 

Higher levels correspond to excitations farther from the boundary.  In
Figs.~\ref{fig:cartoon}(c) and \ref{fig:cartoon}(d) this is illustrated
pictorially by showing the dominant configurations in some of the states
occurring at several levels.

The ES can be understood physically, and constructed accurately and
completely, through the boundary-linked perturbation theory introduced
previously.  Here we present a qualitative discussion highlighting the
algebraic structure of the problem. For the XXZ model, this theory can be
formulated as follows.  The vacuum state has no domain walls.  Excitations in
the ES are obtained by introducing domain walls near the boundary and by
moving them into the bulk.  From explicit perturbative calculations at the
first few orders, we find that the only perturbation terms leading to new ES
levels are those which do one of the following: (1) create a pair of domain
walls by applying ${\mathcal H_{xy}}$ across the boundary:
\begin{equation}
\label{pr-2}
\dots{\sigma}{\sigma}\u\d]
[\u\d{\sigma}{\sigma}\dots\quad\leadsto\quad
\dots{\sigma}{\sigma}\underline{\u\u}]
[\underline{\d\d}{\sigma}{\sigma}\dots  \; ,
\end{equation}
or (2) act twice, on the first bonds on either side of the boundary, again
creating a domain wall pair:
\begin{equation}
\label{pr-1}
\dots{\sigma}\d\u\d]
[\u\d\u{\sigma}\dots\quad\leadsto\quad
\dots{\sigma}\underline{\d\d}\u]
[\d\underline{\u\u}{\sigma}\dots
\end{equation}
or (3) act twice symmetrically with respect to the boundary so as to move a
domain wall on each block, each by two bonds away from the boundary.  Above,
domain wall positions are shown underlined, and $\sigma$ represents a spin of
unspecified orientation.
The first process changes $S_A^z$ by $\pm{1}$, while the others preserve  $S_A^z$. 
The
boundary-locality of the ES is encoded in the restriction that domain walls
are created only through the first two processes above and not farther away
from the boundary.  The symmetrical pairwise application of $\mathcal{H}_{xy}$
ensures that the new contribution is to an element of the $\mathbf{M}$ matrix
that is not on the same row or column as a previous element that already led
to a lower order ES level.

Given a domain wall configuration, the corresponding $A$ state belongs to the
sector ${\delta}S_A^z=(-1)^{N_D}\tfrac{1}{2}\sum_j(-1)^j(1-(-1)^{\ell_j})$,
where $N_D$ is the number of domain walls in $A$, $j\in[1,N_D]$ labels those
starting from the boundary, and $\ell_j \ge1$ is the position of the $j$-th
domain wall measured from the cut.  In our perturbative rules, the order at
which a particular configuration appears is given by $n=\sum_j\ell_j$.  One
can thus construct the ES, organized by ${\delta}S_A^z$ sector, from the
perturbative rules in terms of the domain walls.  In this way one obtains
exactly the ES of Fig.~\ref{fig:cartoon}(a).

For each ${\delta}S_A^z$, the lowest ES level (filled symbols in
Fig.~\ref{fig:cartoon}(a)) is generated by a state where the region near the
boundary is packed with domain walls at each bond, creating a ferromagnetic
region as large as needed to have that ${\delta}S_A^z$ value.  The lowest
level for sector ${\delta}S_A^z$ occurs at order
$m({\delta}S_A^z)=|{\delta}S_A^z(2{\delta}S_A^z-1)|$.

The degeneracies of the ES follow intriguing patterns, which can be explained
through our construction.  The degeneracy at order $n$ in a given
${\delta}S_A^z$ tower is $p([n-m({\delta}S_A^z)]/2)$ where $p(x)$ is the
number of integer partitions of the integer $x$.  The total degeneracy at a
given order $n$, listed to the right of Fig.~\ref{fig:cartoon}(a), is given by
$q(n)$, the number of partitions into unequal summands
\cite{integ_seq_Sloane}.  The degeneracy sequence $q(n)$ was observed in
corner transfer matrix calculations \cite{Peschel-99}, but our perturbative
construction now gives a physical picture: the number of ES levels at order
$n$ is the number of ways one can place domain walls within the first $n$
positions while keeping the sum of position labels ($\sum_j\ell_j$) to be $n$;
this number is $q(n)$ by definition.

\begin{figure}
\includegraphics[width=0.99\columnwidth]{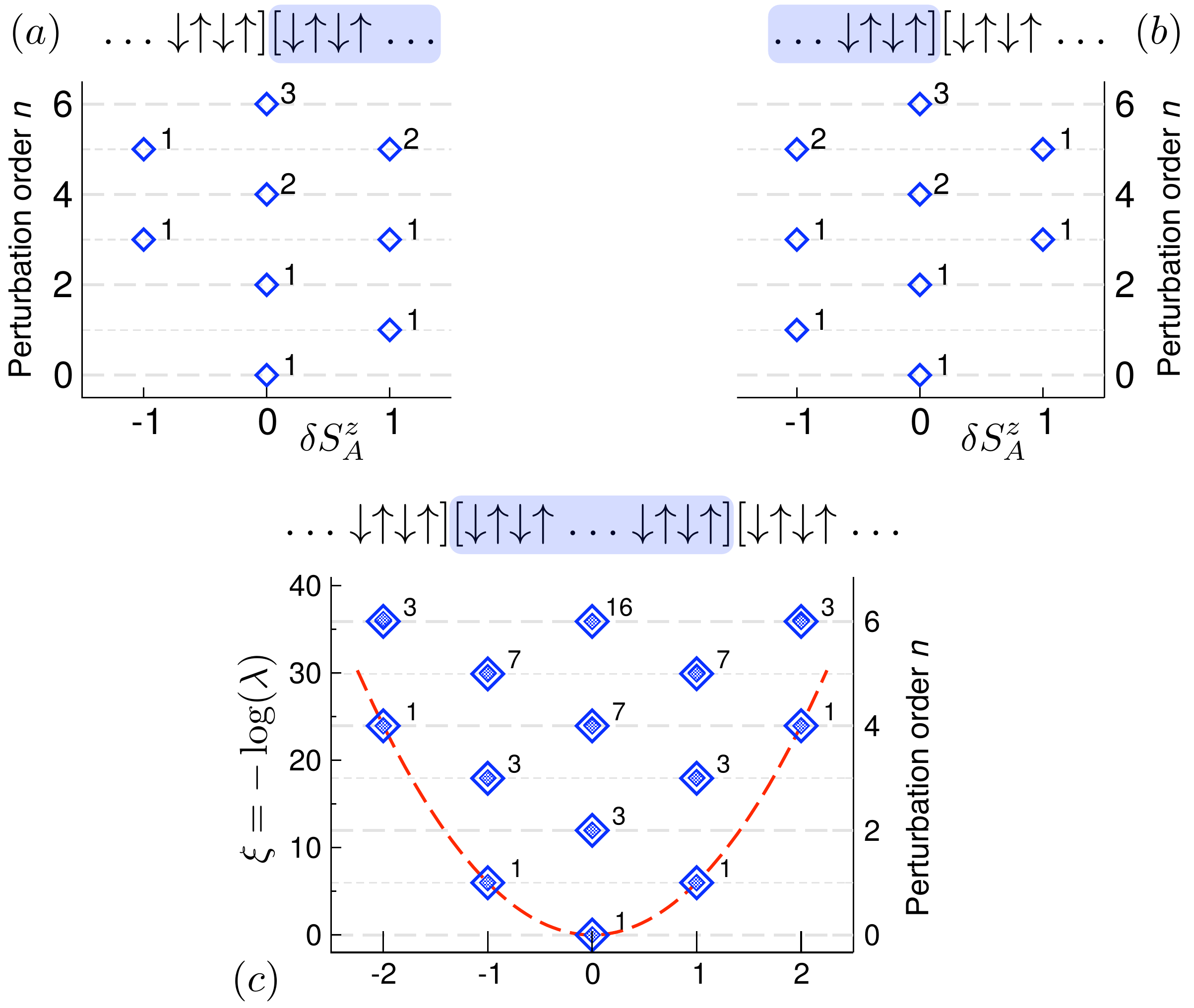}
\caption{(Color online) Combining two single-boundary ES to form a two-boundary ES, for the
  XXZ chain.  Shading indicates the $A$ partition.  (a,b) The single-boundary
  ES obtained by considering two different bipartitions.  (c) Combination of
  single-boundary cases to form a two-boundary ES (large empty diamonds),
  compared with numerical diagonalization data at $\Delta=10$ (small filled
  diamonds), which match perfectly.  }
\label{fig:comb}
\end{figure}

For the infinite XXZ chain, this ES is actually exact for all $\Delta>1$
(not just $\Delta\gg{1}$), with the ES level spacing $2\trm{arccosh}\Delta$,
as a result of integrability \cite{Peschel-99}.  In a generic (non-integrable)
gapped phase, the hierarchical level structure is most pronounced in the
case of weak perburbations, and gets progressively broadened by higher order
renormalizations of entanglement levels upon increasing the perturbation.

Fig.~\ref{fig:cartoon}(e) plots the exact ES for a finite open chain.  
The ES follows the infinite-system structure at low orders, and only deviates
significantly (e.g. split degeneracies) at orders corresponding to the
distance between the boundary and the physical edge.  This illustrates again
that the lowest ES levels correspond to smallest distances from the partition
boundary.

\paragraph*{The two-boundary ES ---}

Boundary-locality implies that the single-boundary ES is sufficient to
construct the ES for multiple-boundary blocks when the boundaries are
sufficiently far apart.  Formally, this physical intuition means that the
reduced density matrix $\rho_A$ of a two-boundary block should `factorize' in
the loose sense that $\rho_A$ is iso-spectral to $\rho_{L}\otimes\rho_{R}$,
where $\rho_{L(R)}$ are (virtual) reduced density matrices appropriate for
single-boundary blocks having the left (right) boundary of $A$. This idea has
recently be shown to be at work for ES of blocks of a fractional quantum Hall state
on the torus~\cite{Laeuchli_PRL10}.

Fig.~\ref{fig:comb} shows the combination of two single-boundary ES to form a
two-boundary ES in an XXZ chain.  The case shown corresponds to the ground
state being a dressed single N\'eel state $\ket{N1}$, and an even size for the
$A$ block.  The relevant single edge ES, Figs.~\ref{fig:comb}(a) and
\ref{fig:comb}(b), are then ($\delta{S_A^z}$)-inverted versions of each other.
Each level of Fig.~\ref{fig:comb}(a) is combined with each level of
Fig.~\ref{fig:comb}(b), the two constituents contributing additively to the
order and to $\delta{S_A^z}$.  The resulting ES has the structure and
degeneracies shown in Fig.~\ref{fig:comb}(c).
In the combined ES, the lowest level at any ${\delta}S_A^z$ is found, from a
combination of $m_L(x)=|x(2x+1)|$ and $m_R(x)=|x(2x-1)|$, to be
$m({\delta}S_A^z)=({\delta}S_A^z)^2$.
This explains the parabolic envelope shown as dashed line in
Fig.~\ref{fig:comb}(c).

If the $A$ block is odd instead of even, the two single-boundary ES are
identical rather than reflections of each other; the combined ES would then
not be symmetric under ${\delta}S_A^z$ inversion.

\begin{figure}
\includegraphics[width=\linewidth]{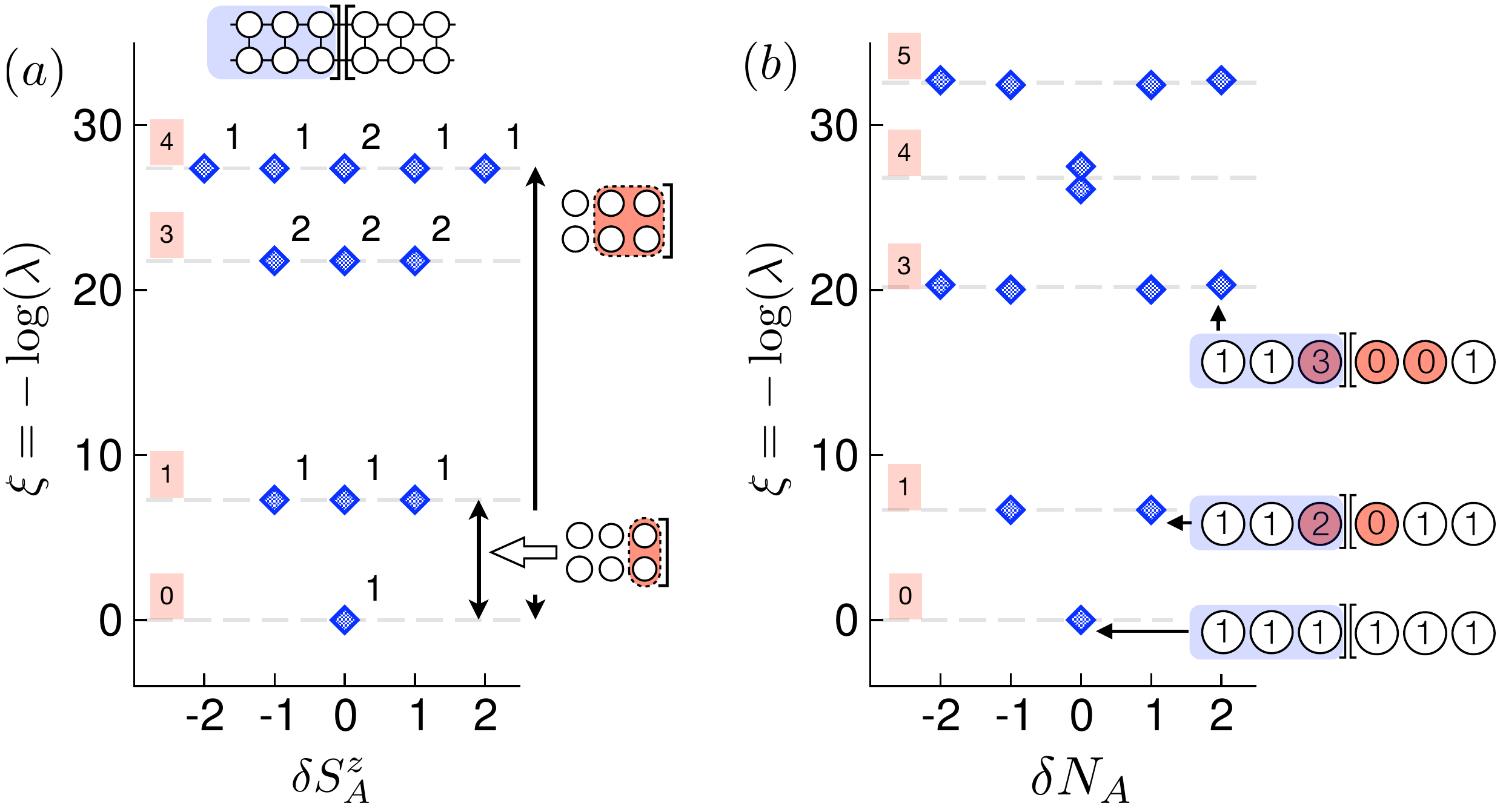}
\caption{(Color online) Single-boundary ES (DMRG data) for the (a) Heisenberg spin ladder and (b)
  Bose-Hubbard chain.  For the ladder, we show numerical data for
  $J_{leg}=0.1J_{rung}$, in the rung singlet phase.
The partitioning is shown in the top cartoon (region $A$ shaded as in Fig.~\ref{fig:cartoon}). 
The integers next to the data points indicate approximate degeneracies.  In the associated cartoons,
we show with shading the rungs (degrees of freedom) which contribute to
various levels of the ES.
For the Bose-Hubbard case (data shown for $U=40$), the
  cartoons  show dominant contributions to ES levels.  Sites with
  occupancy $\neq1$ are shaded; occupancies are indicated with integers
  on each site.   Partition $A$ is again as in Fig.~\ref{fig:cartoon}.
In both panels the horizontal light dashed lines are guides to the eye indicating the perturbation 
order at which the levels appear.
}
\label{fig_ladder}
\end{figure}

\paragraph*{Other gapped 1D systems ---}

Boundary-locality can be illustrated with the ES of various other gapped
systems.
As a first additional example we demonstrate in Fig.~\ref{fig_ladder}(a) the
case of the gapped spin-$\tfrac{1}{2}$ Heisenberg ladder~\cite{Dagotto-96}
with a cut which bisects both chains (in contrast to the chain - chain setup studied
in Refs.~\cite{Poilblanc-10,Cirac-11,Peschel11}).
There is once again a hierarchical level structure corresponding to increasing
distances from the boundary.
The levels have $SU(2)$ structure; there is a triplet at first order, no new
levels at second order, two triplets at third order, a singlet and a
quintuplet at fourth order, etc.  Boundary-linked perturbation arguments can
explain this structure, and also the parabolic envelope.

Our last example system is the  Bose-Hubbard chain in the gapped Mott insulating
phase at unit filling (Fig.~\ref{fig_ladder}b).  The level structures in the ES are  richer
than the XXZ or ladder cases, due to the larger local Hilbert space.  Some
features, like a near-parabolic envelope, appeared in Ref.~\cite{Santos-11}.
Here we provide the physical picture behind this observation based on 
boundary-locality, as illustrated through the cartoons in Fig.~\ref{fig_ladder}.  
The nature of the local Hilbert spaces now allow asymmetric excitations in the two blocks,
leading to a modified parabola for the envelope: 
$\xi \approx x(x+1)\ln{U} -\ln\left[(x+1)!\right] + \mathcal{O}(1/U)$, with $x=|{\delta}N_A|$.

\paragraph*{Conclusions ---}

While entanglement as a boundary-related quantity is a direct reflection of
the area law and is thus well-appreciated, the competing picture
of a `bulk' entanglement Hamiltonian clearly calls for a thorough exploration
of boundary-locality and its consequences.  We have provided the elements of
such an analysis here.  In particular, we have shown how boundary-locality
allows a perturbative formalism specific to calculating entanglement spectra.
At any finite perturbative order, we reproduce a finite number of the lowest
ES levels, which amounts to constructing a finite-rank matrix product
representation of the state.

To put our work in context of recent literature: high temperature series
expansions for the mutual information exhibit a boundary linked
property~\cite{Singh11} similar to our rather different $T=0$ perturbation theory
targeting ES.  Other recent works highlighted the structure of ES along a 1D
block boundary in 2D systems~\cite{qi-2012,Cirac-11}. Our work is complementary
in that we clarify the structure of the ES perpendicular to the cut, i.e. as
one moves further into the bulk.

Our work opens up several research avenues.  For example, in gapless 1D
states, and in Fermi liquids in higher dimensions, the entanglement entropy
does not scale purely with the boundary size but has a logarithmic correction.
This implies some kind of weakening of the boundary-local picture; degrees of
freedom deep in the bulk must play a greater role in such cases.  Quantifying
this effect remains an open task.

\newpage
\section{Supplementary Materials}

In these Supplements, we 
\\ (1) 
present general features of the perturbative structure of the `Schmidt'
($\mathbf{M}$) matrix and the entanglement spectrum (ES), and a relatively
general proof of boundary-locality at low perturbation orders;
\\ (2) provide some details concerning the perturbative
calculation of ES in the Heisenberg spin ladder.


\section{Overview; the `Schmidt' ($\mathbf{M}$) matrix}

The purpose of these Supplementary Materials is to provide some details and
examples of perturbative calculations of the entanglement spectrum (ES), both
in specific and general settings.  The general formulation also provides a
proof of boundary-locality for the ES, for a wide class of gapped 1D systems
described by local Hamiltonians.

It is convenient to describe our calculation in terms of the matrix
$\mathbf{M}$, appearing in the Schmidt decomposition process, as introduced in
the main text:
\[
|\psi\rangle=\sum\limits_{i,j}{\bf
M}_{ij}|\varphi^A_i\rangle\otimes|\varphi^B_j\rangle
\]  
where $|\varphi_i^A\rangle$ and $|\varphi_j^B\rangle$ are two orthonormal
bases for the subsystems $A$ and $B$.  Since the matrix occurs during Schmidt
decomposition, we refer to it as the \emph{Schmidt matrix}.  The singular
value decomposition (SVD) of $\mathbf{M}$ provides the Schmidt coefficients.
The eigenvalues $\{\lambda_i\}$ of the reduced density matrix are the squares
of the singular values of $\mathbf{M}$.  It is very convenient, therefore, to
express perturbative considerations on the ES, $\{\xi_i=-\log\lambda_i\}$, in
terms of the $\mathbf{M}$ matrix.

In the main text, we mentioned the following observation: 

\begin{quote}

When calculating ${\bf M}$ perturbatively, we find through examination of
several examples that a contribution to ${\bf M}$ at some order can lead to a
new ES level only if the contribution is not appearing on the same row or the
same column as a previous contribution which led to a new ES level at lower
order.  In physical terms, this means that we do not get new ES levels by
applying a perturbation only to the $A$ block but need to apply perturbation
terms to both blocks in a linked way.

\end{quote}

This observation is related to how the rank of the $\textbf{M}$ matrix grows
when the matrix $\textbf{M}$ is being built up perturbatively.  These somewhat
abstract statements will be made concrete in these Supplements, by explicitly
writing out the $\textbf{M}$ matrix for several cases.  These demonstrations
of Schmidt matrices will also clarify the physical significance of this
feature of perturbative ES calculations, in terms of distinguishing between
`bulk' and `boundary-connected' contributions to the $\textbf{M}$ matrix.

Section \ref{sec_general} will present the perturbative structure of the
$\mathbf{M}$ matrix in a very general form, demonstrating the above statement
(and additional features of $\mathbf{M}$) under relatively mild assumptions,
without reference to an explicit Hamiltonian.  This analysis will also lead to
a general `proof' of boundary-locality of the ES, for low perturbation orders.
This complements the material in the main text, where our physical results on
boundary-locality have been illustrated mainly through examples on explicit
physical systems and not generically.

Section \ref{sec_ladder} will present the perturbative calculation of the ES
for the case of the Heisenberg ladder in some detail, for first and second
order.  This complements the main text where the main example was the XXZ
chain, and where we focused on physical concepts at the expense of detailed
formulas.  In particular, we will show for this explicit model how the
$\textbf{M}$ matrix grows perturbatively, leading to more and more ES levels
at increasing orders.

\section{Perturbative structure of ES and of $\mathbf{M}$ matrix: general features} \label{sec_general}

In this section we present in very general form the perturbative structure of
the ES and of the $\mathbf{M}$ matrix, without relying on a particular example
system and using only mild assumptions about the form of the Hamiltonian.
This also leads to a relatively general proof of boundary-locality, for low
perturbation orders.
We do not attempt or claim completeness of proof.  Our purpose is to show why
boundary-locality is expected to hold as a general feature of the ES in gapped
phases connected to a product state.

We consider the system Hamiltonian to be of the form
\begin{equation}
\label{ham}
{\mathcal H} ~=~ \sum_i{\mathcal H}_i
~=~ \sum_i{\mathcal H}^0_i ~+~ \epsilon\sum\limits_{i}
{\mathcal H}'_i
\end{equation}
where $\epsilon$ is a small parameter.  The interaction is a sum of local
terms, which we assume to be finite range but not necessarily single-site.
The hamiltonian is translationally invariant (${\mathcal H}_i= {\mathcal
H}_{i+1}$ $\forall i$), except at the system edges. 

The unperturbed ground state is taken to be a product state, as is true in
all our examples.  The unperturbed Hamiltonian ($\sum_i{\mathcal H}^0_i$) may
be either single-site or finite-range --- in our example systems, ${\mathcal
H}^0_i$ are single-site (single-rung) terms in the Bose-Hubbard chain and the
Heisenberg ladder, but of nearest-neighbor form in the XXZ chain.

We specialize to the generic case of a nearest neighbor interaction of the
form ${\mathcal H}'_i=K^1_{i}K^2_{i+1}+h.c.$, where $K^1_i$, $K^2_i$ are
arbitrary operators acting on site $i$.  This includes a large class of
models, including the Bose-Hubbard chain and the XXZ chain.  It is
straightforward but cumbersome to generalize our results below to the case of
next-neighrest-neighbor (or generic finite range) interactions, or to other
local Hamiltonians.

For simplicity, we write below only contributions arising from the $K^1_{i}
K^2_{i+1}$ term, and omit the terms or elements arising from the hermitian
conjugate. (Or, equivalently, we specialize to systems where $K^1_{i}
K^2_{i+1}$ is hermitian.)  Again, including these contributions would be
straightforward but cumbersome, and does not change any of the general
conclusions below.  For simplicity we will also assume that applying $K^1_{i}
K^2_{i+1}$ on the unperturbed ground state any number of times generates
product states which are also eigenstates of the unperturbed Hamiltonian
$\sum_i{\mathcal H}^0_i$.

The unperturbed product state $|0\rangle=\prod_i|0_i\rangle$ has the simple
Schmidt decomposition  $|0\rangle\equiv |0\rangle_A\otimes|0\rangle_B$.
Therefore, at order zero in perturbation theory, the $\mathbf{M}$ matrix has
only the single element $a_{11}=1$.

\paragraph*{\underline{First order} ---}

For first order perturbation theory, we can choose as basis for the $A$
subsystem the states
\begin{eqnarray*}
&|0\rangle_{A}&
\\
& c_1\, K^1_iK^2_{i+1}|0\rangle_{A}&  \qquad i,i+1\in A
\\
&|0'\rangle_{A}&
\end{eqnarray*}
and a corresponding basis set for the $B$ block.  
There are as many `bulk' states $K^1_iK^2_{i+1}|0\rangle_{A}$ in this basis as
there are bonds in $A$.  ($c_1$ is the normalization.)
The two ``boundary states'' $|0'\rangle_{A(B)}$ are
defined such that 
\begin{align*}
K^1_j K^2_{j+1}|0\rangle_A\otimes|0\rangle_B = c_2
|0'\rangle_A\otimes|0'\rangle_B \\ \mathrm{with}
\quad j\in A \, , \quad j+1\in B
\end{align*}
i.e., $|0'\rangle_{A(B)}$ are the $A$($B$) states produced when the
perturbation $H'_j = K^1_j K^2_{j+1}$ acts across the boundary.
We thus have a total of $1+(L_{A(B)}-1)+1$ basis states, where $L_{A(B)}-1$ is
the number of bonds in block $A$($B$).  

[Our omission of the hermitian conjugate has simplified the number of
``boundary terms'' at first order.  For a generic non-factorizable interaction
of the form $K_{j,j+1}$ we would have $K_{j,j+1}|0_j\rangle\otimes
|0_{j+1}\rangle=\sum_{m,n} c_{m
n}|\alpha_j^m\rangle\otimes|\alpha_{j+1}^n\rangle$ where
$\alpha_j^m$ is a basis for the local Hilbert space at site
$j$.]

With the above-described ($1+[L_{A(B)}-1]+1$)-dimensional bases for $A$($B$)
blocks, the ${\bf M}$ matrix at first order is of $(L_B+1)\times(L_A+1)$ form
\[
\begin{pmatrix}
a_{11} & \tilde{a}\epsilon & \tilde{a}\epsilon & \cdots & \tilde{a}\epsilon & 0\\
\tilde{a}\epsilon & 0 & 0 &\cdots  & 0 & 0\\
\tilde{a}\epsilon & 0 & 0 & \cdots & 0 & 0\\
\vdots & \vdots &\vdots & & 0 & \vdots\\
\tilde{a}\epsilon & 0 & 0 &  \cdots & 0 & 0\\
0 & 0 & 0 &  \cdots & 0 & a_{33}\epsilon
\end{pmatrix}  \, .
\]
This matrix can be simplified by choosing the more convenient basis for the
$A$ block
\[
|0\rangle_{A} \, , \quad \quad \frac{c_1}{\sqrt{L_{A}-1}}\sum_{i,i+1\in A}
K^1_iK^2_{i+1}|0\rangle_{A} \, , \quad \quad |0'\rangle_{A}
\]   
and the corresponding set for block $B$.  The idea here is that the
$K^1_iK^2_{i+1}|0\rangle_{A}$ states can be combined together as they all
provide the same contribution.  Going to this new basis first for the $A$
block and then for the $B$ block, we get the `shrinking'
\begin{widetext}
\[
\begin{pmatrix}
a_{11} & \tilde{a}\epsilon & \tilde{a}\epsilon & \cdots & 0\\
\tilde{a}\epsilon & 0 & 0 &\cdots  & 0\\
\tilde{a}\epsilon & 0 & 0 & \cdots & 0\\
\vdots & \vdots &\vdots & & \vdots\\
0 & 0 & 0 &  \cdots &a_{33}\epsilon
\end{pmatrix}
~\longrightarrow~
\begin{pmatrix}
a_{11} & a_{12}\epsilon & 0\\
\tilde{a}\epsilon & 0 & 0\\
\tilde{a}\epsilon & 0 & 0\\
\vdots & \vdots &\vdots\\
0 &  \cdots &a_{33}\epsilon
\end{pmatrix}
~\longrightarrow~
\begin{pmatrix}
a_{11} & a_{12}\epsilon & 0\\
a_{21}\epsilon & 0 & 0\\
0 & 0 & a_{33}\epsilon
\end{pmatrix} 
\]
\end{widetext}
so that the  ${\bf M}$ matrix can be written as 
\[
\mathbf{M} ~=~ \begin{pmatrix}
a_{11} & a_{12}\epsilon & 0\\
a_{21}\epsilon & 0 & 0 \\
0 & 0 & a_{33}\epsilon
\end{pmatrix}
\]
with $a_{12}\equiv \tilde{a}\sqrt{(L_A-1)}$ and $a_{21}\equiv \tilde{a}\sqrt{(L_B-1)}$.
This ``shrinking'' property holds in general: if  ${\bf M}$ has an
isolated block with only one nonzero row or column, since its rank is
trivially one it is possible to shrink the block to form a column or a row
with only one nonzero element.

The matrix ${\bf M}$ written above has a $2\times2$ block with the two
Schmidt (singular) values:  
\begin{eqnarray}
\label{first}
a_{11}+\frac{a_{12}^2+a_{21}^2}{2a_{11}}\epsilon^2\\
\label{second}
\frac{a_{12}a_{21}}{a_{11}}\epsilon^2
\end{eqnarray}

In a first-order calculation, $O(\epsilon^2)$ contributions to the Schmidt
values are not meaningful.  The second singular value above is thus `fake', and
the $2\times2$ block has rank 1, up to meaningful order.  The $\ord(\epsilon^2)$
correction to the first singular value is also meaningless.

Thus, the $a_{12},a_{21}$ appearing on the first row and column respectively
do not give rise to any first order Schmidt value since they contribute only
at order $\ord(\epsilon^2)$.  This illustrates in a general setting, not
referring to a particular model, that matrix elements appearing on the same
row or column as previous contributions which gave lower order ES levels do
not give new levels.  Such terms only contribute to a renormalization of
$|0\rangle_{A(B)}$.  We note that these non-contributing terms originate from
perturbative processes acting only on one of the two subsystems.

In contrast, the element $a_{33}\epsilon$ forms a separate block by itself,
and thus contributes a new $\ord(\epsilon)$ ES level.  This is an example of
an element appearing perturbatively in a new row and new column of the
$\mathbf{M}$ matrix and giving rise to a new ES level.  This new ES level
comes from the boundary terms $|0'\rangle_{A(B)}$.

We thus have proved in a very general setting that, at first order, bulk terms
disjoint from the boundary do not contribute to the ES and the only
contribution comes from the boundary.

\paragraph*{\underline{Second order} --- }

At second order, we need to include extra basis states.  All or most of these
new states are formed by applying ${\mathcal H}'_i$ twice within the same
block.  This leads to a set of $\ord(\epsilon^2)$ contributions which in the
${\bf M}$ matrix fall on the same row or column as $a_{11}$.  As in first
order, we can shrink these to a single new row and a single new column, and as
in first order we can show that they do not provide any true ES contributions
at this order.  Again, bulk contributions are seen not to contribute to the ES
at low order.  We omit such terms in the expressions written below.

More interesting second-order contributions are those where ${\mathcal H}'_i$
acts once in $A$ and once in $B$.  This can be written within the basis set
used previously for first order.  The $2\times2$ block of the  $\mathbf{M}$
matrix, discussed previously at first order, takes the form 
\[
\begin{pmatrix}
a_{11} & a_{12}\epsilon \\
a_{21}\epsilon & a_{22}\epsilon^2 
\end{pmatrix}
\]
The singular values from this $2\times2$ block are
\begin{eqnarray}
a_{11}+\frac{a_{12}^2+a_{21}^2}{2a_{11}}\epsilon^2\\
\frac{a_{12}a_{21}-a_{22}a_{11}}{a_{11}}\epsilon^2
\end{eqnarray}
The extra value (as well as the renormalization to $a_{11}=1$) is now relevant
since it is of same order as the perturbation order.  We thus can have a new
contribution, but only if $a_{22}\neq{a}_{12}a_{21}$.  

This condition again reflects boundary locality.  For second-order terms where
the two applications $H'_j$ occur away from the boundary in $A$ and $B$, and
therefore away from each other, the contributions `factorize' into two
first-order applications of $H'_j$ (encoded in the $a_{12}$, $a_{21}$ terms).
The physical processes that might cause $a_{22}\neq{a}_{12}a_{21}$ are
therefore applications of the perturbation connected to the boundary.  
(This intuitive argument is difficult to prove in full generality, but we have
found it to hold in all the systems we have treated.)
In our example systems, we also find that such a second-order contribution
also requires the unperturbed Hamiltonian (${\mathcal H}^0_i$) to be of longer
range than single-site or single-rung terms.  Thus the XXZ chain has such
second order ES levels, while the Heisenberg ladder and the Bose-Hubbard chain
lacks such second order levels, as can be seen from Figures 1 and 3 in the main
text.

There are additional possibilities at second order, but they result in minor
variations of the points discussed above.

Thus, also at second order in perturbation theory, contributions to the ES are
seen very generally to come only from boundary-connected processes.

\bigskip  \bigskip


\section{Heisenberg Spin ladder} \label{sec_ladder}

In this section, we describe perturbative calculations of the ES in the
spin-$\hf$ ladder with $SU(2)$ Heisenberg interactions.  We have performed
calculations up to third order, but the expressions and notation become
extremely cumbersome to write out, so we give details up to second order.  All
the physical points we want to demonstrate are already visible by this stage
of the perturbative calculation.

\subsection{Hamiltonian}

\begin{figure}
\centerline{\includegraphics[width=0.95\linewidth]{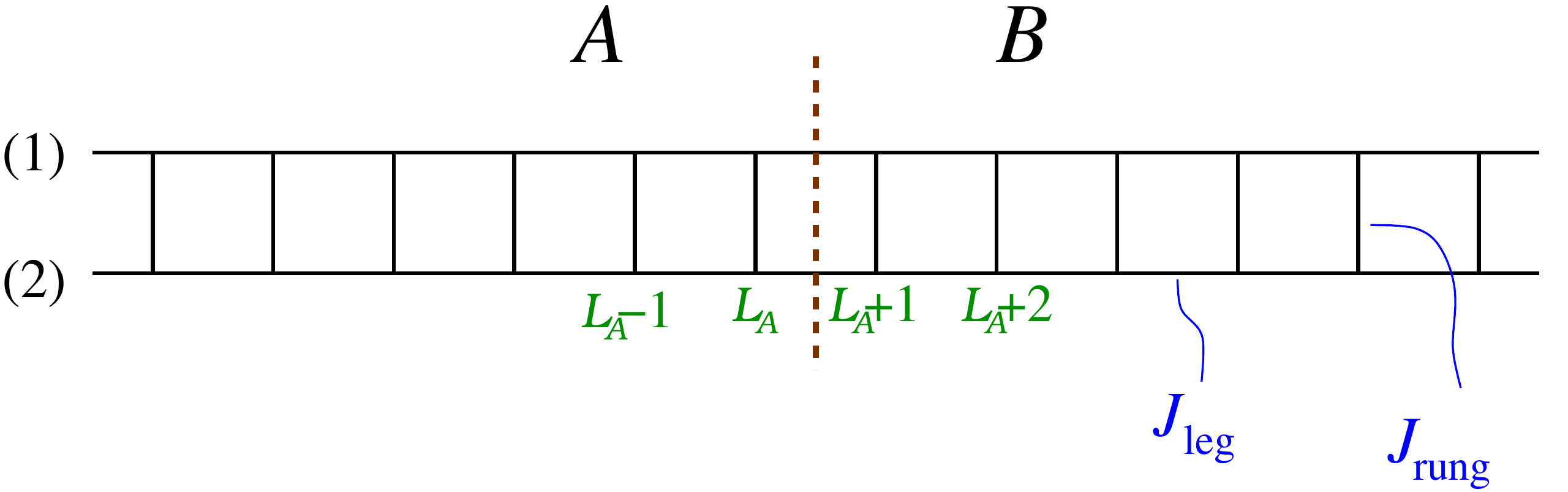}}
\caption{ Spin ladder.  The boundary between $A$ and $B$ regions is shown with
a dashed vertical line.  The rungs in block $A$ and $B$ are numbered from $1$
to $L_A$ and from $L_A+1$ to $L_A+L_B$, respectively.
\label{fig:cartoon_ladder}}
\end{figure}

The Hamiltonian for the two-leg Heisenberg ladder is
\begin{eqnarray}
H ~&=&~ H_{rung} +H_{leg}  \nonumber  \\
~&=&~ J_{rung} \sum\limits_{i}S_i^{(1)}\cdot S_{i}^{(2)} \nonumber \\
&&~+~  J_{leg}\sum\limits_{i}\bigg(S_i^{(1)}\cdot S_{i+1}^{(1)}+
S_i^{(2)}\cdot S_{i+1}^{(2)}\bigg)
\end{eqnarray}
where $J_{rung}$ is the coupling between spins on the same rung, and $J_{leg}$
is the coupling along each leg (Figure \ref{fig:cartoon_ladder}).  The
subscript $i$ indexes rungs, and (1),(2) refer to the two legs.  Open boundary
conditions are used, so that there is a single boundary between the blocks,
shown with a dashed line in Figure \ref{fig:cartoon_ladder}.  The number of
rungs is $L_A$ in block $A$ and $L_B$ in block $B$; total $L=L_A+L_B$.

We are interested in the limit $J_{leg}\ll J_{rung}$ where $H_{leg}$ can be
treated as a perturbation.  We will use $J_{leg}=1$.  The small parameter for
perturbation is
\[
\epsilon = J_{leg} / J_{rung} = 1 /  J_{rung} \, .
\]
The eigenstates of the rung Hamiltonian $H_{rung}$ are conveniently expressed
in terms of the single rung states. These are the singlet $|s\rangle$ and the
three triplets $|t^\alpha\rangle$:
\begin{gather*}
|s\rangle\equiv\frac{1}{\sqrt{2}}(|\uparrow\downarrow\rangle-|\downarrow
\uparrow\rangle) 
\; ; \quad \quad 
|t^0\rangle\equiv\frac{1}{\sqrt{2}}(|\uparrow\downarrow\rangle+|\downarrow\uparrow\rangle)
\; ;
\\
|t^+\rangle\equiv|\uparrow\uparrow\rangle \; ; \quad\quad
||t^-\rangle\equiv|\downarrow\downarrow\rangle 
\; .
\end{gather*}
The rung energy is $-\frac{3}{4}J_{rung} = -\frac{3}{4\epsilon}$ for the
singlet and $\frac{1}{4} J_{rung} = \frac{1}{4\epsilon}$ for the triplets.
The difference is $1/\epsilon$. 

The zeroth order ground state is the product state
\[
\psi^{(0)} = \ket{ss\cdots ss}\equiv\ket{s}_1\otimes\ket{s}_2\cdots\otimes\ket{s}_L
\]
i.e., each rung is a singlet state. Other unperturbed eigenstates are those in which 
some of the rungs have been converted to $\ket{t^{\alpha}}$.  

To consider the result of applying $H_{leg}$ on any state, it is convenient to
know how $H_{leg}$ acts on a pair of rungs:
\begin{align*}
H_{leg}|ss\rangle=\frac{1}{2}\sum\limits_{\alpha}\zeta^\alpha|t^\alpha 
t^{\bar\alpha}\rangle\\
H_{leg}|t^0t^0\rangle=\frac{1}{2}(|ss\rangle +|t^+t^-\rangle +|t^- 
t^+\rangle)\\
H_{leg}|t^+t^-\rangle=\frac{1}{2}(-|t^+t^-\rangle+|t^0t^0\rangle-|ss\rangle)\\
H_{leg}|t^-t^+\rangle=\frac{1}{2}(-|t^-t^+\rangle+|t^0t^0\rangle-|ss\rangle)\\
H_{leg}|t^\alpha s\rangle=\frac{1}{2}|st^\alpha\rangle\quad \alpha=+,-,0 \\
H_{leg}|t^\beta t^0\rangle=\frac{1}{2}|t^0t^\beta\rangle\quad \beta=+,-
\end{align*}
In the first row the sum is over $\alpha=+,0,-$ and the sign function $\zeta^\alpha$ is
$\zeta^\alpha=-1$ if $\alpha=+,-$ and $\zeta^\alpha=1$ for $\alpha=0$.

\subsection{First order}

A single application of the perturbation operator creates a
$\ket{t^{\alpha}}\ket{t^{\bar{\alpha}}}$ pair in the sea of $\ket{s}$ rungs.
The wavefunction at first order is
\begin{gather*}
\psi^{(1)} = \psi^{(0)}  +~  \sum_{i\alpha} c_{i\alpha} \psi_{i\alpha}^{(0)} 
\\
\mathrm{with} \quad  \quad 
c_{i\alpha} = -\frac{\bra{\psi_{i\alpha}^{(0)}
  }H_{leg}\ket{\psi^{(0)}}}{2(1/\epsilon)} = - \zeta^{\alpha} \epsilon/4 
\; .
\end{gather*}
Here $\psi_{i\alpha}^{(0)}$ is the state where the singlet pair $i$, $i+1$ has
been promoted to a $\ket{t^{\alpha}t^{\bar{\alpha}}}$ pair.  The unperturbed
ground state $\psi^{(0)}$ appears without a subscript.
\begin{multline*}
\psi^{(1)} = \Big( \ket{s}\cdots\ket{s}\Big) \\ ~-~ \frac{\epsilon}{4} \sum_{i} 
\Big( \ket{s}_1\cdots\ket{t^{0}t^0}_{i}
\cdots\ket{s}_L \Big)  \\
 ~+~ \frac{\epsilon}{4} \sum_{i} \Big( \ket{s}_1\cdots\ket{t^+t^-}_{i}\cdots
\ket{s}_L \Big)   \\
 ~+~ \frac{\epsilon}{4} \sum_{i} \Big( \ket{s}_1\cdots\ket{t^{-}t^{+}}_{i}
\cdots\ket{s}_L \Big)
\;  .
\end{multline*}
We introduce the following concise notation for states: all sites which are in
$\ket{s}$ states are omitted and only the $\ket{t^{\alpha}}$'s are explicitly
written, with the subscript indicating the site or the bond.   Thus
\begin{align}
\label{first_or}
\psi^{(1)}=\psi^{(0)}  -\frac{\epsilon}{4} \sum_\alpha\sum\limits_{i}
\zeta^\alpha\ket{t^{\alpha}t^{\bar\alpha}}_{i}  \; .
\end{align}
Note that the norm deviates from unity by some $\ord(\epsilon^2)$ value, i.e.,
is correct at first order.

We will now construct the Schmidt ($\mathbf{M}$) matrix at first order.  The block spin
component $S_A^z$ is a good quantum number; therefore we treat each spin
sector separately.

\paragraph*{\underline{$S_A^z=1$ sector} ---}

We start with the $S_A^z=1$ sector.  The relevant basis for the two subsystems
$A$ and $B$ is
\begin{equation}
{\mathcal B}_A=\{\ket{t^+}_{L_A}\} \, , \qquad {\mathcal
B}_B=\{\ket{t^-}_{L_A+1}\} \, .
\end{equation}
The triplets are on rungs neighboring the boundary between the blocks.  The
$\ket{s}$ rungs have been omitted.

The Schmidt matrix in this sector has a single element, $\epsilon/4$, which
gives the reduced density matrix eigenvalue $\lambda' =\epsilon^2/16$, and the
entanglement level
\[
\xi = -\log[(\epsilon/4)^2]=-2\log(\epsilon/4) \, . 
\]
The same result is obtained in the $S_A^z=-1$ sector.

\paragraph*{\underline{$S_A^z=0$ sector} ---}

The $S_A^z=0$ sector is more interesting.  We choose the block bases
\begin{eqnarray*}
{\mathcal B_A}&=&\{\ket{s},\, \ket{\phi_A},\, \ket{t^0}_{L_A}\}\\
{\mathcal B_B}&=&\{\ket{s},\, \ket{\phi_B},\, \ket{t^0}_{L_A+1}\}
\end{eqnarray*}
where $\ket{s}$ denotes the unperturbed state of that subsystem, and we defined
\begin{eqnarray}
\ket{\phi_{A/B}}=\frac{1}{\NN_{A/B}}\sum
\limits_{\alpha,i}
\zeta^\alpha\ket{t^\alpha t^{\bar\alpha}}_i 
\end{eqnarray}
where $\NN_{A/B}= \sqrt{3(L_{A/B}-1)}$ is the norm.  As in
Section \ref{sec_general} we are combining states from boundary-disconnected
excitations into a single state, to `shrink' rows and columns of the
$\textbf{M}$ matrix.  With this choice the $\textbf{M}$ matrix in this sector
has the form
\[
\begin{pmatrix}
1 & -\frac{\epsilon}{4}\NN_A &  0\\
-\frac{\epsilon}{4}\NN_B & 0 & 0\\
0 & 0 &  -\frac{\epsilon}{4}  
\end{pmatrix} 
\; .
\]
The isolated element $-\epsilon/4$ gives the same contribution to the ES as
the $S_A^z=\pm 1$ sectors, completing the $S=1$ multiplet at
$\xi=-2\log(\epsilon/4)$.  (Lowest triplet in Figure 3a of the main text).

The remaining $2\times2$ block has two singular values, but one of them is of
higher than linear order in $\epsilon$, and is thus a `fake' singular value
that does not contribute to this order of the perturbative calculation.  In
other words, the effective rank of the $2\times2$ block, at the order being
considered, is 1.

Specifically, the reduced density matrix eigenvalues from this block are
\begin{eqnarray}
\lambda_0 &=& 1+\frac{1}{16}(\NN_A^2+\NN_B^2)\epsilon^2 + \ord(\epsilon^4)
\label{fir_ord_lambdas_plus}
\\
\lambda_1 &=& \frac{1}{256}(\NN_A\NN_B)^2\epsilon^4 + \ord(\epsilon^6) 
\label{fir_ord_lambdas_minus}
\end{eqnarray}
The second ($\lambda_1$) is `fake', as explained above.

Thus, we see here one example of the observation quoted in the Introduction:
since the first-order contributions $\epsilon\NN_{A,B}$ appear on the same row
or same column as the zeroth order element 1, they do not lead to new ES
levels.  On the other hand, the isolated matrix element $-\epsilon/4$ appears on a
new row and new column, and therefore is allowed to give rise to a new ES
level, which it does.

Note that the `fake' eigenvalues are size-dependent: this reflects the physics
that they arise from perturbations acting on the bulk and not those acting on
an edge-connected location.  To neglect terms like $\epsilon^2\NN_{A}\NN_{B}$ in
the thermodynamic limit, one has to use the correct order of limits.

The size-dependent correction in $\lambda_0$ to its unperturbed value (unity)
is also `fake'.  We will later see that this fake correction gets canceled at
next order in the perturbation.


\subsection{2nd order}

The state up to second order can be written as
\begin{multline}
\psi^{(2)}=\Big(1-\frac{3}{32}\epsilon^2(L_A+L_B-1)\Big)\psi^{(0)}  \\
- \bigg(\frac{\epsilon}{4}+\frac{\epsilon^2}{8}\bigg)\sum\limits_{\alpha,i} 
\zeta^{\alpha}\ket{t^\alpha t^{\bar\alpha}}_i +
\frac{\epsilon^2}{8}\sum\limits_{\alpha,i} \zeta^{\alpha}\ket{t^\alpha
  st^{\bar\alpha}}_i  
\\+\frac{\epsilon^2}{16}\sum\limits_{\alpha,\beta,i} \sum\limits_{n}  \zeta^\alpha \zeta^\beta
|t^\alpha t^{\bar\alpha} 
s^nt^{\beta}t^{\bar\beta}\rangle_i
\end{multline}
We use the notation for total wavefunctions introduced in the previous
subsection.  In addition, here $s^n$ denotes a sequence of $n$ rungs in state
$s$.  

\paragraph*{\underline{$S_A^z=1$ sector} ---}

We start again with the $S_A^z=1$ sector, which at first order had the lone
element $\epsilon/4$. We choose the following
basis 
\[
{\mathcal B}_A=\{\ket{t^+}_{L_A},\ket{\phi'_A}\} \, , \quad
{\mathcal B}_B=\{\ket{t^-}_{L_A+1},\ket{\phi'_B}\} \, ,
\]
where we defined
\begin{equation}
\ket{\phi'_A}\equiv\frac{1}{\NN'_A}\bigg(\ket{t^+s}+\frac{1}{2}\sum\limits_{\alpha,
  n} \zeta^\alpha |t^\alpha t^{\bar\alpha} s^nt^+\rangle\bigg)
\end{equation}
with $\NN'_A=\sqrt{1+\frac{3}{4}(L_A-2)}$, and analogously for the $B$ block.  When we omit a site subscript on a block
ket, the symbols in the ket refer to the configurations of the rungs ending at
the boundary.  The rungs further from the boundary are of course understood to
be singlets.

In this basis the Schmidt matrix is
\[
\begin{pmatrix}
(\frac{\epsilon}{4}+\frac{\epsilon^2}{8}) & & +\frac{\epsilon^2 \NN'_A}{8}\\\\
+\frac{\epsilon^2 \NN'_B}{8} & & 0
\end{pmatrix}  \, ,
\]
from which the reduced density matrix eigenvalues are found to be
\begin{eqnarray}
\lambda' &=& \frac{\epsilon^2}{16}+\frac{\epsilon^3}{16}+\frac{1}{64}(1+\NN_A^2+\NN_B^2)
\epsilon^4+O(\epsilon^5) \\
\lambda'' &=& \frac{1}{256}(\NN_A\NN_B)^2(\epsilon^6)+O(\epsilon^7) \, .
\end{eqnarray}
We see that $\lambda'$ has acquired an $\ord(\epsilon^3)$ correction, while
the size-dependent $\ord(\epsilon^4)$ term is spurious.  The $\lambda''$ is a
completely spurious eigenvalue; the meaningful rank of this $2\times2$ block
is 1, at this order.

\paragraph*{\underline{$S_A^z=0$ sector} ---}

The $S_A^z=0$ sector itself has the block structure
\[
\begin{pmatrix}
M_s & 0 \\ 0 & M_t
\end{pmatrix}
\]
where the triplet part $M_t$ is the same as the matrix obtained above for the
$S_A^z={\pm}1$ case.

For the singlet part $M_s$, we use the  $A$ block basis 
\begin{eqnarray}
{\mathcal B}_A=\{|s\rangle, |\phi_A\rangle, |\phi''_A\rangle\}
\end{eqnarray}
where
\begin{multline*}
\ket{\phi''_A}\equiv\frac{1}{\NN''_A}
\Bigg(
\sum_{\alpha,i}
\zeta^\alpha\ket{t^\alpha st^{\bar\alpha}}_i  ~+~
\\ \frac{1}{2}
\sum\limits_{\alpha,\beta,i}\sum\limits_{n} \zeta^\alpha \zeta^\beta
|t^\alpha t^{\bar\alpha} s^nt^\beta t^{\bar\beta}\rangle_i
\Bigg)
\end{multline*}
with $\NN''_A=\sqrt{3(L_A-2)+\frac{9}{8}(L_A-2)(L_A-3)}$.
We find the singlet part of the $S_A^z=0$ Schmidt matrix to be 
\[
M_s=
\begin{pmatrix}
1-\frac{\epsilon^2}{32}(\NN_A^2+\NN_B^2+3) & &
-(\frac{\epsilon}{4}+\frac{\epsilon^2}{8})\NN_A & & \frac{\epsilon^2}{8}\NN''_A\\\\
-(\frac{\epsilon}{4}+\frac{\epsilon^2}{8})\NN_B & &\frac{\epsilon^2}{16}\NN_A\NN_B & & 0\\\\
\frac{\epsilon^2}{8}\NN''_B & & 0 & & 0 
\end{pmatrix}
\, .
\]
The term $\epsilon^2(\NN_A^2+\NN_B^2+3)/32$ appears due to normalization.  The
reduced density matrix eigenvalues obtained from $M_s$ are, for simplicity,
written below only for the case $L_A=L_B$:
\begin{equation}
\lambda_0=1-\frac{3}{16}\epsilon^2+\frac{3}{8}(L_A-1)\epsilon^3+O(\epsilon^4)
\end{equation}
\begin{multline}
\lambda_{\pm} =\frac{9}{8192}(L_A-1)\bigg(-14+9L_A+3L_A^2 
\\ \pm 4\sqrt{2(L_A-1)(3L_A^2+L_A-6)}\bigg)\epsilon^6 
\end{multline}
The extensive $\ord(\epsilon^2)$ term that appeared in
Eq.~\eqref{fir_ord_lambdas_plus} is no longer present in $\lambda_0$, and the
contribution $3\epsilon^2/16$ is genuine second order.  The spurious
eigenvalue $\lambda_1$ of Eq.~\eqref{fir_ord_lambdas_minus} has been pushed
from $\ord(\epsilon^4)$ to $\ord(\epsilon^6)$ in the $\lambda_{\pm}$ here.

\begin{figure}[b]
\centerline{\includegraphics[width=0.9\linewidth]{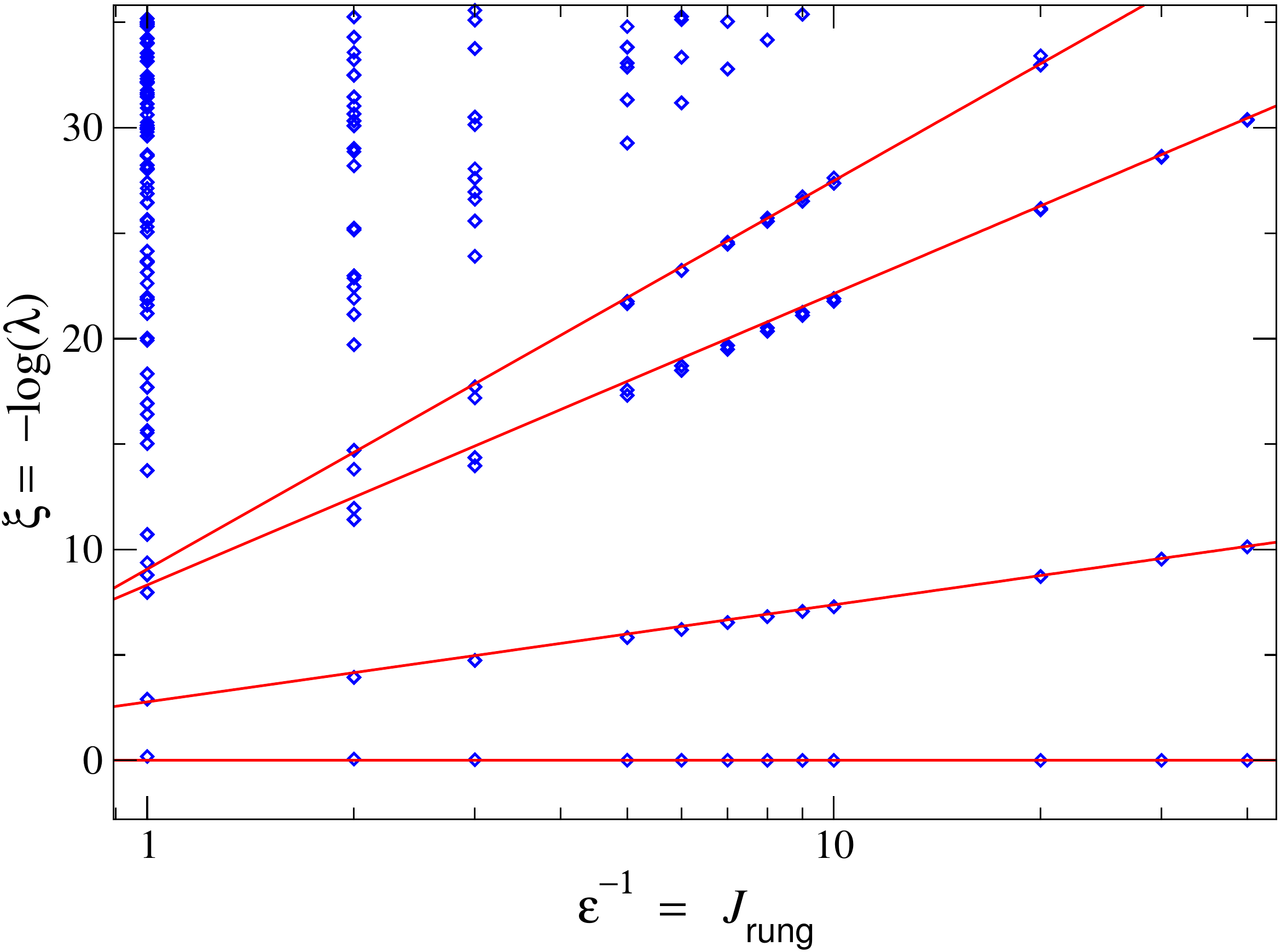}}
\caption{ Ladder entanglement spectrum, obtained using DMRG for a 20-rung
ladder ($L_A=L_B=10$), as a function of
$J_{rung}=\epsilon^{-1}$.  The perturbative expressions are shown as red
straight lines.
\label{fig:ES_ladder_ED_Jrung}}
\end{figure}

\bigskip

\subsection{Perturbative expressions for ES levels}

We have elaborated on the perturbative calculations at the first and second
orders.  We do not show details of the next orders, but the perturbative
expressions at the lowest few orders for the ES levels, and the leading orders
of the corresponding entanglement eigenstates, are listed below:
\begin{align*}
\begin{array}{rllll}
&\xi_0=0   & |ss\rangle\\\\
&\xi_1=-2\log(\frac{\epsilon}{4})   
    & |st^\alpha\rangle\\\\
&\xi_2=-2\log(\frac{\epsilon^3}{64})   
    & \Big\{\frac{1}{\sqrt{2}}\big(|t^\beta t^0\rangle-|t^0t^\beta\rangle\big),\\
& & \frac{1}{\sqrt{2}}\big(|t^+ t^-\rangle-|t^-t^+\rangle\big),\\ 
    & &|t^\alpha s\rangle\Big\}\\\\
&\xi_3=-2\log(\frac{11}{1024}\epsilon^4)  
    & \Big\{\frac{1}{\sqrt{2}}\big(|t^\beta t^0\rangle+|t^0t^\beta\rangle\big),\\
    & &\frac{1}{\sqrt{6}}\big(|t^+ t^-\rangle+|t^-t^+\rangle\big)+\\ 
    & & \sqrt{\frac{2}{3}}|t^0t^0\rangle\Big\}
\end{array}
\end{align*}
where $\beta=\pm$ while $\alpha=\pm,0$.    The notation for the eigenvectors
above is that only the states of the two rungs of the $A$ block nearest the
boundary are shown.  

A noteworthy feature is that there is no $\ord(\epsilon^2)$ contribution.
This reflects an aspect of boundary-locality on which we have commented in
detail in Section \ref{sec_general}.

The perturbative expressions for ES levels are shown as red straight lines in
Figure \ref{fig:ES_ladder_ED_Jrung}, and are seen to describe the exact ES
levels rather well at low orders.
\end{document}